\documentstyle[12pt]{article}
\title{\bf Noncommutativity in quantum cosmology and the Hierarchy problem}
\author{F. Darabi \thanks{e-mail:
f.darabi@azaruniv.edu}, A. Rezaei-Aghdam \thanks{e-mail:
rezaei-a@azaruniv.edu}, A. R. Rastkar
\\
{\small Department of Physics, Azarbaijan University of Tarbiat
Moallem, 53714-161, Tabriz, Iran .}}
\begin{document}
\maketitle
\begin{abstract}
We study the quantum cosmology of an empty (4+1)-dimensional
Kaluza-Klein cosmology with a negative cosmological constant and a
FRW type metric with two scale factors, one for 4-D universe and
the other for one compact extra dimension. By assuming the
noncommutativity in the corresponding mini-superspace we suggest a
solution for the Hierarchy problem, at the level of
Wheeler-DeWitt equation. 
\end{abstract}

\newpage

Recently, the study of various physical theories from
noncommutative point of view, such as string theory \cite{Seib},
quantum field theory (see for instance \cite{Szaz}), quantum
mechanics \cite{Chai}, and classical mechanics \cite{Romer}, has
been of particular interest. In particular, a new interest has
been developed to study the noncommutative {\it quantum cosmology}
\cite{Garc}, \cite{Barb1}, \cite{Barb2}. In these studies, the
influence of noncommutativity at early universe was explored by
the formulation of a version of noncommutative quantum cosmology
in which a deformation of {\it mini-superspace} is required
instead of space-time deformation \cite{Garc}.

We know that in the study of homogeneous universes, the metric
depends only on the time parameter. Thus, one can find a model
with a finite dimensional configuration space, the so-called
mini-superspace, whose variables are the three-metric components.
The quantization of these models can be performed by using of the
rules of quantum mechanics. The mini-superspace construction is a
procedure to define quantum cosmological models in the search for
describing the quantum features of the early universe. The
influence of noncommutativity on the mini-superspace has already
been considered for the cosmological models with Kantowski-Sachs
and FRW metrics \cite{Garc}, \cite{Barb1}. Here, we consider the
effect of noncommutativity on the configuration space in the model
which was previously studied by one of the authors \cite{Dar}.
This model shows an empty (4+1) dimensional Kaluza-Klein universe
with a negative cosmological constant and a FRW type metric having
two scale factors. Following the idea in regarding the
Wheeler-Dewitt equation more
primitive than the classical Einstien equations \footnote{%
This idea was followed by Hawking and Page investigating the
wormholes \cite{HP}} we study the noncommutativity in the quantum
cosmology of this model and try to propose a solution for the
Hierarchy problem, in the context of quantum cosmology.

The experimental upper bound on the value of cosmological constant
is extremely small. On the other hand, it is usually assumed that
an {\it effective} cosmological constant describes the energy
density of the vacuum $<\rho_{vac}>$. In fact, the vacuum energy
density $<\rho_{vac}>$ is a quantum field theory contribution to
the effective cosmological constant
$$
\Lambda_{eff}=\lambda+ \kappa <\rho_{vac}>,
$$
where $\lambda$ is a small bare cosmological constant. The
calculations show that, these contributions affect enormously the
value of effective cosmological constant as $<\rho_{vac}> \sim
M_P^4$, where $M_P$ is the Planck mass which defines the
ultraviolet cutoff scale of the quantum gravity. This is the
well-known cosmological constant problem. However, there is
another fundamental energy scale in nature, namely the electroweak
scale $M_{EW}$ whose experimental investigation is of particular
interest and the corresponding interaction has been probed
successfully. Over the past two decades there was a great interest
to explain the smallness of $M_{EW}/M_P$, which is known as the
Hierarchy problem. Recently, a great amount of interest has been
concentrated on solving this problem based on the existence of
large extra dimensions \cite{Hamed}. Some attempts have also been
done to solve this problem based on the noncommutativity in the
space-time coordinates \cite{Wang}. To the authors knowledge, this
problem has not yet been paid attention by considering the
noncommutativity in the mini-superspace coordinates of a quantum
cosmology.

In this paper, we introduce a mechanism based on the existence of
noncommutativity in the mini-superspace of a quantum cosmology
corresponding to an empty $(4+1)$-dimensional Kaluza-Klein
cosmology. It is to be noted that the present mechanism does not
claim to solve the problem in a fundamental way because we just
solve the problem in a special and simple model. However, this may
shed light on the similar approaches to solve the problem in a
more fundamental way.

We start with the metric considered in \cite{Dar} in which the
space-time is assumed to be of FRW type which has a compact space,
namely the circle $S^1$. We adopt the chart $\{t, r^i, \rho\}$
with $t$, $r^i$ and $\rho$ denoting the cosmic time, the space
coordinates and the compact space coordinate, respectively. We
therefore take
\begin{equation}
d s^2 = -{dt}^2+R^2 (t) \frac{dr^i \: dr^i}{(1+\frac{k
r^2}{4})^2}+a^2 (t) {d\rho}^2, \label{2}
\end{equation}
where $k=0,\pm 1$ and $R(t)$, $a(t)$ are the scale factors of the
universe and compact dimension, respectively. The curvature scalar
corresponding to metric (\ref{2}) is obtained as
\begin{equation}
{\cal R}=6
\left[\frac{\ddot{R}}{R}+\frac{k+{\dot{R}}^2}{R^2}\right] +2\frac{
\ddot{a}}{a}+ 6 \frac{\dot{R}}{R} \frac{\dot{a}}{a},
\label{3}
\end{equation}
where a dot represents differentiation with respect to $t$.
Substituting this result into Einstein-Hilbert action with a
cosmological constant $ \Lambda$
\begin{equation}
I=\int\!\sqrt{-g} ({\cal R}-\Lambda) dt\:d^3 r\:d\rho,
\label{4}
\end{equation}
and integrating over spatial dimensions gives an effective
Lagrangian $L$ in the mini-superspace ($R$, $a$) as
\begin{equation}
L=\frac{1}{2} R a {\dot{R}}^2+\frac{1}{2}R^2
\dot{R}\dot{a}-\frac{1 }{2}k R a+\frac{1}{6} \Lambda R^3 a.
\label{5}
\end{equation}
By defining $\omega ^2 \equiv -\frac{2\Lambda }3$ and changing the
variables as
\begin{equation}
u=\frac 1{\sqrt{8}}\left[ R^2+Ra-\frac{3k}\Lambda \right] ,
\hspace{10mm}v=\frac 1{\sqrt{8}}\left[ R^2-Ra-\frac{3k}\Lambda
\right],
\label{6}
\end{equation}
$L$ takes on the form
\begin{equation}
L=\frac{1}{2} \left[({\dot u}^2-\omega ^2u^2)-({\dot v}^2-\omega
^2v^2)\right].
\label{7}
\end{equation}
The assumption that the full (4+1) dimensional Einstein equations
 hold, implies that the Hamiltonian corresponding to $L$ in
(\ref{7}) must vanish, that is
\begin{equation}
H=\frac{1}{2} \left[({\dot u}^2+\omega ^2u^2)-({\dot
v}^2+\omega^2v^2)\right]=0,
\label{8}
\end{equation}
which describes an isotropic oscillator-ghost-oscillator system.

The corresponding quantum cosmology is described by the
Wheeler-DeWitt equation resulting from Hamiltonian (\ref{8}) and
can be written as
\begin{equation}
\{ [p_u^2+\omega^2u^2]-[p_v^2+\omega^2v^2] \}\Psi(u, v)=0.
\label{9}
\end{equation}
Now, we consider the effect of noncommutativity on the
configuration space in the above model. The noncommutative quantum
mechanics is defined by the following commutators \cite{Chai}
\begin{equation}
[u,v]= i\theta, \:\:\:\: [u,p_u]= i, \:\:\:\: [v,p_v]= i, \:\:\:\:
[p_u,p_v]= 0, \label{12}
\end{equation}
where we use the natural units ($\hbar=1$). The corresponding
noncommutative Wheeler-DeWitt equation can be written by use of
the star product as \cite{Chai}, \cite{Garc}
\begin{equation}
H*\Psi=0. \label{13}
\end{equation}
We can always represent the noncommutative harmonic oscillator in
terms of anisotropic harmonic oscillator in the commuting
coordinates \cite{Chai}
\begin{equation}
[u,v]= 0, \:\:\:\: [u,p_u]= i, \:\:\:\: [v,p_v]= i, \:\:\:\:
[p_u,p_v]= 0. \label{14}
\end{equation}
Therefore, we obtain anisotropic oscillator-ghost-oscillator
Wheeler-DeWitt equation
\begin{equation}
\left \{ \left[\frac{{p_u}^2}{4}+\omega^2(u-\frac{\theta
p_v}{2})^2\right]-\left[\frac{{p_v}^2}{4}+\omega^2(v+\frac{\theta
p_u}{2})^2\right] \right \}\Psi(u, v)=0, \label{15}
\end{equation}
where the transformations $u\rightarrow u - \theta p_v$ and $v
\rightarrow v + \theta p_u$ have been used. However, in two
dimensions we can also represent the two-dimensional anisotropic
oscillator-ghost-oscillator as the two-dimensional isotropic
oscillator-ghost-oscillator in the presence of an effective
magnetic field as follows
\begin{equation}
\{[(p_u-A_u)^2+{\omega^\prime}^2u^2]-[(p_v-A_v)^2+{\omega^\prime}^2v^2]
\}\Psi(u, v)=0, \label{17}
\end{equation}
where
\begin{equation}
\left \{
\begin{array}{ll}
{\omega^\prime}^2\equiv \frac{4\omega^2}{(1-\omega^2\theta^2)^2},
\\
\\
A_u \equiv \frac{2\omega^2\theta }{1-\omega^2\theta^2}v,
\\
\\
A_v \equiv -\frac{2\omega^2\theta }{1-\omega^2\theta^2}u.
\end{array}\right.
\label{16}
\end{equation}
The above equation can be written in a more convenient form
\begin{equation}
\left \{
\left[p_u^2+({\omega^\prime}^2-\frac{B^2}{4})u^2\right]-\left[p_v^2+({\omega^\prime}^2-\frac{B^2}{4})
v^2\right]+B(vp_u+up_v)\right \}\Psi(u, v)=0, \label{17}
\end{equation}
where $B=-\frac{4\omega^2\theta }{1-\omega^2\theta^2}$ is derived
through ${\bf B}={\bf \nabla}\times {\bf A}$ from
\begin{equation}
A_u= -\frac{B}{2}v \:\:\:,\:\:\: A_v= \frac{B}{2}u. \label{18}
\end{equation}
The $B$-term in Eq.(\ref{17}) deserves more scrutiny so that we
can find the correct frequency for this system. To this end, it is
useful to compare this oscillator-ghost-oscillator system
(\ref{17}) with the corresponding oscillator-oscillator system. By
straightforward calculations we find for the latter system
\begin{equation}
H=\frac{1}{2}\sum_{i=1,
2}[(p_i-A_i)^2+\omega^2x_i^2]=\frac{1}{2}\sum_{i=1,
2}[p_i^2+(\omega^2+\frac{B^2}{4})x_i^2]-\frac{1}{2}B(x_1p_2-x_2p_1),
\label{17'}
\end{equation}
where $A=(-x_2, x_1, 0)\frac{B}{2}$. The $B$-term in
Eq.(\ref{17'}) is the magnetic potential energy and
$\sqrt{(\omega^2+\frac{B^2}{4})}$ is the effective frequency of
the system. It is obvious that the $B$-term has an independent
role and does not contribute to the oscillator frequency, at all.
Now, comparing the two systems we realize that each term in
Eq.(\ref{17'}) has a corresponding term in Eq.(\ref{17}). For
example, the first two brackets in Eq.(\ref{17'}) correspond to
the first two brackets in Eq.(\ref{17}), and the third $B$-term in
Eq.(\ref{17'}) corresponds to the third $B$-term in Eq.(\ref{17}).
Therefore, the corresponding terms have the same roles with the
difference that the terms in Eq.(\ref{17}) have the {\it ghost}
characters. This occurs in other examples, as well. For instance,
the two-dimensional oscillator with no external magnetic field is
defined by
\begin{equation}
H=(p_x^2+p_y^2)+\omega^2(x^2+y^2), \label{+}
\end{equation}
whereas the corresponding oscillator-ghost-oscillator system is
defined by
\begin{equation}
H=(p_x^2-p_y^2)+\omega^2(x^2-y^2). \label{-}
\end{equation}
Here, the first parenthesis in both systems has the kinetic term
role and the second parenthesis plays the role of potential term,
with the difference that the terms in Eq.(\ref{-}) have the ghost
characters.

Therefore, as the $B$-term in Eq.(\ref{17'}) is an independent
potential term which does not contribute to the effective
frequency $\sqrt{(\omega^2+\frac{B^2}{4})}$, the corresponding
$B$-term in Eq.(\ref{17}) should also be an independent potential
term which does not contribute to the effective frequency, as
well.

In this regard, we can write
\begin{equation}
\tilde{\omega}^2={\omega^\prime}^2-\frac{B^2}{4}=\frac{4\omega^2}{(1-\omega^2\theta^2)},\:\:\:
\omega^2\theta^2\leq 1, \label{19}
\end{equation}
which defines the resultant frequency of the isotropic
oscillator-ghost-oscillator in the presence of a constant magnetic
field $B$. This frequency comes from a ghost-like combination of
the oscillator term ${\omega^\prime}^2$ and the cyclotron term
$\frac{B^2}{4}$.

In Eq.(\ref{9}), the oscillator frequency $\omega$ was defined by
the effective cosmological constant
$\Lambda_{eff}=-\frac{3}{2}\omega^2$. Comparing with
Eq.(\ref{17}), we suppose $\tilde{\omega}$ to be defined by a new
effective cosmological constant
$\tilde{\Lambda}_{eff}=-\frac{3}{2}\tilde{\omega}^2$. Substituting
these definitions into Eq.(\ref{19}) we obtain
\begin{equation}
\tilde{\Lambda}_{eff}=
\frac{4\Lambda_{eff}}{(1-\frac{2}{3}\theta^2|\Lambda_{eff}|)}.
\label{20}
\end{equation}
This is a redefinition of the effective cosmological constant due
to the noncommutativity. However, we know that the effective
cosmological constant, in principle, is a measure of the
ultraviolet cutoff in the theory. Therefore, the above equation
can also be considered as a redefinition of the cutoff in the
theory due to the noncommutativity. This is the main result which
is considered to solve the Hierarchy problem in the present model.
To this end, we take
\begin{equation}
\theta^2=\frac{3}{2}\frac{M_P^4-4M_{EW}^4}{M_P^4M_{EW}^4},
\label{21}
\end{equation}
where $M_{EW}$ is the electroweak mass scale. Moreover, we take
$\Lambda_{eff}\sim M_{EW}^4$ representing $M_{EW}$ as the natural
cutoff in the original commutative model. Therefore, we obtain
from the above equation
\begin{equation}
\tilde{\Lambda}_{eff}\sim M_P^4,
\label{22}
\end{equation}
which defines the cutoff in the noncommutative model. In other
words, if we assume $M_{EW}$ to be the natural cutoff in the
original commutative model, the Planck mass is then the cutoff in
the noncommutative model. This solves the Hierarchy problem at the
level of quantum Wheeler-DeWitt equation by assuming that $M_{EW}$
is the only fundamental mass scale in the model, and $M_P$ is the
mass scale which is appeared due to introducing the
noncommutativity in the mini-superspace. Put another way, one may
suppose that the universe, in principle, has just one fundamental
energy scale for all interactions, namely $M_{EW}$. The quantum
gravity sector of this universe must then be described by the
Wheeler-DeWitt equation (\ref{9}) with the vacuum energy density
of the same scale $\omega^2=-\frac{2}{3}\Lambda_{eff}\sim
M_{EW}^4$. However, the energy scale of the quantum gravity which
we expect to experience in the universe is defined by the Planck
mass $M_P$ and not $M_{EW}$. This discrepancy between $M_P$ and
$M_{EW}$, namely the Hierarchy problem is solved by the assumption
of noncommutativity in the quantum gravity sector of the universe
which leads Eq.(\ref{9}) to Eq.(\ref{17}) with the vacuum energy
density of the Planck scale
$\tilde{\omega}^2=-\frac{2}{3}\tilde{\Lambda}_{eff}\sim M_{P}^4$.
It then turns out that $M_P$ is not a new fundamental scale and
its enormity $M_P\gg M_{EW}$ is simply a consequence of the
noncommutativity in the quantum gravity sector of the model,
namely the Wheeler-DeWitt equation.

It is worth noting that having the exact numeric coefficients in
Eq.(\ref{21}) is very important to get Eq.(\ref{22}) which solves
the Hierarchy problem. This in particular means that our proposed
solution for the Hierarchy problem comes with a {\it fine
tunning}. Although this fine tunning makes the proposed solution
less appealing, however, it is still a novel method, as the
Hierarchy problem seems to be solved with tunning only one
parameter, $\theta$.

\section*{Acknowledgment}

We would like to thank the referee very much for the useful and
constructive comments. This work has been financially supported by
the Research Department of Azarbaijan University of Tarbiat
Moallem, Tabriz, Iran.


\begin{thebibliography}{99}
\bibitem{Seib}N. Seiberg and E. Witten, JHEP {\bf 09} (1999), 032.
\bibitem{Szaz}R. Szazbo, Phys. Rep {\bf 378} (2003), 207.
\bibitem{Chai}M. Chaichian, M. M. Sheikh-Jabbari, and A. Tureanu,
Eur. Phys. J. C {\bf 36} (2004) 251; L. Mezincescu, {\it Star
operation in quantum mechanics}, hep-th/0007046; J. Gamboa, M.
Loewe, and J. c. Rojas, Phys. Rev. D {\bf 64} (2001), 067901; S.
Bellucci and A. Nersessian, Phys. Lett. B {\bf 542} (2002) 295.
\bibitem{Romer}J. M. Romero, J. A. Satiago, and D. Vergara, Phys. Lett. A {\bf 310} (2003),
9; A. E. F. Djemai, {\it On noncommutative classical mechanics},
hep-th/0309034.
\bibitem{Garc}H. Garc\'{i}a-Compe\'{a}n, O. Obreg\'{o}n, and C.
Ram\'{i}rez, Phys. Rev. Lett {\bf 88} (2002), 161301.
\bibitem{Barb1}G. D. Barbosa and N. Pinto-Neto, Phys. Rev. D{\bf 70} (2004), 103512.
\bibitem{Barb2}G. D. Barbosa, Phys. Rev. D{\bf 71} (2005), 063511.
\bibitem{Dar}F. Darabi and H. R. Sepangi, Class. Quantum. Grav
{\bf 16} (1999), 1656.
\bibitem{Hamed}N. Arkani-Hamed, S. Dimopoulos, and G. Dvali, Phys. Lett. B{\bf 429}
(1998), 263; L. Randall, R. Sundrum, Phys. Rev. Lett.{\bf 83}
(1999), 3370.
\bibitem{Wang}F. Lizzi, G. Mangano, and G. Miele, Mod. Phys. Lett. A{\bf 16} (2001),
1; Xiao-Jun Wang, {\it Cosmological Constant as Vacuum Energy
Density of Quantum Field Theories on Noncommutative Spacetime},
hep-th/0411248.
\bibitem{HP}S. W. Hawking, D. N. Page, Phys. Rev. D{\bf 42} (1990), 2655.
\end{thebibliography}
\end{document}